\documentclass[12pt,a4paper]{article}
\usepackage{graphicx}
\usepackage[a4paper,left=1cm,right=1cm]{geometry}
\usepackage[latin1]{inputenc}
\usepackage{amsmath}
\usepackage{amsfonts}
\usepackage{amssymb}
\usepackage{braket}
\usepackage{amsthm}
\usepackage{bbm}
\usepackage{color}
\usepackage{dsfont}
\usepackage{slashed}
\usepackage{hyperref}
\usepackage{cite}

\def\be{\begin{equation}}
\def\ee{\end{equation}}
\def\bea{\begin{eqnarray}}
\def\eea{\end{eqnarray}}

\newcommand{\e}{{\mathrm e}}
\newcommand{\dd}{{\mathrm d}}
\newcommand{\del}{\partial}
\usepackage[all]{xy}
\makeindex
\begin{document}
\newtheorem{ev}{Everything}[section]
\newtheorem{theorem}[ev]{Theorem}
\newtheorem{lemma}[ev]{Lemma}
\newtheorem{cor}[ev]{Corollary}
\newtheorem{prop}[ev]{Proposition}
\theoremstyle{definition}
\newtheorem{exa}[ev]{Example}
\newtheorem{mydef}[ev]{Definition}
\theoremstyle{remark}
\newtheorem{rema}[ev]{Remark}
\makeatletter
\newtheoremstyle{indented}
  {1pt}% space before
  {1pt}% space after
  {\addtolength{\@totalleftmargin}{1 em}
   \addtolength{\linewidth}{-4 em}
   \parshape 1  1.5 em \linewidth}% body font
  {}% indent
  {\bfseries}% header font
  {.}% punctuation
  {.5em}% after theorem header
  {}% header specification (empty for default)
\makeatother
 \theoremstyle{indented}
\newtheorem{iprop}[ev]{Proposition} 
\newtheorem{ith}[ev]{Theorem} 

\renewcommand{\today}{}
\newcommand{\ii}{ i}

\numberwithin{equation}{section}

%\newcolumntype{M}[1]{>{\centering\arraybackslash}m{#1}}

\title{\vspace{-2cm} The Weyl-Mellin quantization map}

\author{{\Large Alessandro Carotenuto$^{a}$\footnote{acaroten91@gmail.com}, Fedele Lizzi$^{b,c,d}$\footnote{fedele.lizzi@na.infn.it},}\vspace{8pt}
\\{\Large 
Flavio Mercati$^{b,c}$\footnote{flavio.mercati@gmail.com}, Mattia Manfredonia$^{b}$\footnote{mattia.manfredonia91@gmail.com}}
\vspace{12pt}
\\
$^{a}$Institute of Mathematics of the Czech Academy of Sciences;
\vspace{6pt}\\
$^{b}$ Dipartimento di Fisica ``Ettore Pancini'',
\vspace{6pt}\\
Universit\`{a} di Napoli {\sl Federico~II}, Napoli, Italy;
\vspace{6pt}
\\
$^{c}$ INFN, Sezione di Napoli;
%\\
%Complesso Universitario di Monte S. Angelo,
%\\
%Via Cintia Edificio 6, 80126 Napoli, Italy;
\vspace{6pt}
\\
$^{d}$ Departament de F\'{\i}sica Qu\`antica i Astrof\'{\i}sica\\
and Institut de C\'{\i}encies del Cosmos (ICCUB),\\
Universitat de Barcelona, Barcelona, Spain.
}

\maketitle

\vspace{-12pt}

\begin{abstract}
We present a quantization of the functions of spacetime, i.e.\ a map, analog to Weyl map, which reproduces the $\kappa$-Minkowski commutation relations, and it has the desirable properties of mapping square integrable functions into Hilbert-Schmidt operators, as well as real functions into symmetric operators. The map is based on Mellin transform on radial and time coordinates. The map also define a deformed $*$ product which we discuss with examples.
\end{abstract}

\newpage

\section{Introduction}

Dirac, Weyl and Wigner \cite{Weyl,Wigner,Moyal} introduced the procedure of \emph{quantization}, which associate an operator to a function of classical phase space.  The starting point is the standard commutation relation $[p,q]=\ii\hbar$. The procedure has been widely generalised and is very fruitful for both physics and mathematics.  The inverse procedure, where a classical function is associated to a quantum operator, is called \emph{dequantization}. The most common quantization scheme is associated to the name of Weyl, and the inverse process to that of Wigner. There are other schemes and we will comment on them later.

According to a widespread belief, the small scale coarseness of spacetime should manifest itself through a set of noncommuting operators which  play the role of the quantum analogue of coordinates of the classical phase space. This is the point of view of noncommutative geometry, where the elements of a noncommutative algebra are interpreted as functions over a quantum space~\cite{Connes_book94, Landi:1997sh, GraciaBondia:2001tr}.
Among the proposed models of noncommutative spacetimes, the one known as $\kappa$-Minkowski ~\cite{Lukierski:1991ff,Lukierski:1992dt,Majid:1994cy}  is of particular interest. The algebra of this quantum spacetime is generated by the non commutative coordinates $\{ x_0;  x_i\}_{i=1,2,3}$ with commutations relations given by 
\begin{equation}\label{kmcr}
[ x_0, x_i]=\ii \frac1\kappa x_i,
\end{equation}
where the real number $\kappa$ has the dimension of inverse length and sets a scale.

Although the noncommutative geometry of these models has been widely  investigated both in the physical and mathematical literature~\cite{Kosinski1999,Kosinski2001,KowalskiGlikman:2003we,Freidel:2007hk,Arzano:2017uuh,Agostini:2002yd,Arzano:2014jfa,Arzano2007,Arzano:2009ci,Matassa:2013,Matassa:2014,Arzano2018,Gubitosi:2013rna,Mercati:2011aa,Mercati:2018hlc,Mercati:2018ruw,Mercati:2018fba,Ballesteros:2019bnc,Lizzi:2019wto,Amelino-Camelia2007,Moyalarea,Gravityquantumspacetime,Lizzi:2020,Meljanac:2016jwk,Loret:2016jrg,Juric2015}, not much has been said about how it relates to the problem of quantization. 
%by which we mean a procedure to relate functions on a classical phase space, usually $\mathbb{R}^{2N},$ with quantum operators acting on a suitable Hilbert space.

In the present paper we will consider Dirac and Weyl's approach, 
i.e.\ the association of operators to classical quantities. We will consider, as is done for ordinary quantum
mechanics,  different orderings. In particular we present a proposal of a Weyl-like quantization scheme that associates quantum observables over $\kappa$-Minkowski spacetime to classical quantities.
Recently~\cite{Lizzi:2018qaf,Lizzi:2019wto}, three of us have shown that the Mellin transform plays the same role of the Fourier transform in ordinary quantum mechanics for the $\kappa$-Minkowski, since it provides the unitary transformation of the Hilbert space that allows to pass from the eigenbasis of the position operator to the eigenbasis of the time operator. This suggests to consider a quantization procedure \`a la Weyl based on the Mellin transform rather then the Fourier transform. At the same time the algebra of operators obtained through this procedure is generated by the commutation relations~\eqref{kmcr} rather than the canonical commutation relations. Implicit in a quantization procedure there is a notion of star product, and we will discuss the product coming form our procedure. The first attempt to describe the algebra of the $\kappa$-Minkowski space through a star-product was~\cite{Agostini:2005mf}. In~\cite{Meljanac:2006ui,Meljanac:2007xb} a family of star products written as a formal power series was considered.  Other star-product realizations of $\kappa$-Minkowski are~\cite{Dabrowski:2010im,Dabrowski:2010yk} and~\cite{Durhuus:2011ci,Mercati:2011pv,Iochum:2011iu}.
More recently, a notable approach was that of~\cite{Pachol:2015qia,Kupriyanov:2020axe}.

The aim of this paper is to work in analogy with the well-established theory of Weyl quantizations. In particular we characterise the quantizers which send real functions over the classical space into symmetric operators, hence associating (up to the choice of a suitable self-adjoint extension) classical observables with quantum observables. Then we deal with the spectral properties of these objects, showing that square integrable functions are sent into Hilbert-Schmidt operators.  Finally we compute the dequantizers that provide the inverse Wigner maps with respect to the quantization procedures that we are studying. The combination quantizer-dequantizer defines a $*$ product which we discuss in the last section.

\section{Quantization and dequantization schemes.}
 Weyl's quantization map provides the standard rule of quantization in classical quantum mechanics, i.e.\ the one defined by the commutator $[\hat q,\hat p]=i$,  and it is given by:
\begin{align}\label{WM1}
\Omega^{\mathrm{Weyl}}:\; &\mathcal{F}(\mathbb{R}^{2N})\rightarrow Op(\mathcal{H})
\nonumber\\ &f(q,p)\mapsto \hat{F}(\hat{q},\hat{p})=\frac1{2\pi}\int_{\mathbb{R}^{2N}} \dd \eta \dd\omega\tilde{f}(\eta,\omega)\e^{\ii \left(\eta \hat{q}- \omega \hat{p}\right)}.
\end{align}
Here $\mathcal{F}(\mathbb{R}^{2N})$ denotes a suitable class of functions over the classical phase space $\mathbb{R}^{2N}$ and  $\tilde{f}(\eta,\omega)$ is the  (symplectic) Fourier transform of $f(q,p).$ %\textcolor{red}{Il $\frac1{2\pi}$ \`e parte del quantizer o della misura?}
Where it not for the hats over $p$ and $q$ the expression for $F$ would just be the original function. The procedure is based on the presence of the operator $\e^{\ii \left(\eta \hat{q}- \omega \hat{p}\right)}$, which in this case is a function of $(\eta,\omega)$. We call this operator the \emph{Weyl quantizer}:
\begin{align}%\label{QW}
\hat{Q}^{\mathrm{Weyl}}_{\mathcal{F}}:\; &  \mathbb{R}^{2N}\rightarrow Op(\mathcal{H})
\nonumber\\&(\eta, \omega)\mapsto \e^{\ii \left(\eta \hat{q}- \omega \hat{p}\right)}.
\end{align}
More generally, in a space of coordinates $z$,   we call quantizer $\hat{Q}: \mathbb R^{n}\rightarrow Op(\mathcal{H})$  an operator-valued function and the map (up to normalization constant which could be absorbed by the measure)
 \begin{align}
\Omega:  & \mathcal{F}(\mathbb R^{n})\rightarrow Op(\mathcal{H})\nonumber
\\&f(z)\mapsto \int_{\mathbb R^{n}} \dd \mu(z) f(z) \hat{Q}(z)
\end{align}
 the \emph{quantization scheme associated to the quantizer} $ \hat{Q}(z).$

 \begin{rema}: Sometimes it is not sufficient to consider operator-valued functions, which we will use  to define a quantization scheme. For sake of simplicity, we will mostly think to the quantizers as operator valued-functions and we are going to explicitly specify when they are given by distributions.\end{rema}

In the Weyl map defined by~\eqref{WM1}, the Fourier transform appears under the integral, rather than the function $f(q,p)$ itself. In this case in fact the variable $z=(\eta,\omega)$.
This is just a convention, the quantizer may as well be expressed in term of  $p$ and $q$. It suffices to rexpress $\tilde f(\eta,\omega)$ as the symplectic Fourier transform of $f(q,p)$ to obatain:
\be
 \hat{F}(\hat{q},\hat{p})=\frac1{2\pi}\int_{\mathbb{R}^{2N}}\dd q \dd p\, \dd \eta \dd\omega 
 f(q,p)\e^{-\ii \left(\eta {q}- \omega {p}\right)}\e^{\ii \left(\eta \hat{q}- \omega \hat{p}\right)}.
\ee
From which one can read that 
\be
 \hat{F}(\hat{q},\hat{p})=\frac1{2\pi}\int_{\mathbb{R}^{2N}}\dd q \dd p\, 
 f(q,p)\hat Q(q,p)
\ee
with
\be
\hat Q(q,p)= \dd \eta \dd\omega 
 \e^{-\ii \left(\eta ({q}-\hat q) -\omega ({p}-\hat p)\right)}.
\ee

More generally, given $\hat{Q}$ be a quantizer, if there exists another quantizer (in Fourier transform space) $\hat{Q}_{\mathcal{F}}$ such that, for every $f$ with Fourier transform $\tilde{f},$ one has
\begin{equation}
\hat{Q}(f(q,p))=\hat{Q}_{\mathcal{F}}(\tilde{f}(\eta,\omega))
\end{equation}
we call $\hat{Q}_{\mathcal{F}}$ the \textit{Fourier transform} of $\hat{Q}.$

To this extent first use the Campbell-Baker-Hausdorff formula to express the quantizer as the product of two exponentials:
\begin{equation}\label{orderWeyl}
\hat{Q}^{\mathrm{Weyl}}_{\mathcal{F}}(\eta,\omega)=\e^{\ii \left(\eta \hat{q}- \omega \hat{p}\right)}=\e^{\ii \eta \hat{q}}\e^{-  i\omega \hat{p}}\e^{\frac i2 \omega\eta}=\e^{-  i\omega \hat{p}}\e^{\ii \eta \hat{q}}\e^{-\frac i2 \omega\eta}
\end{equation}
Notice that 
\begin{equation}
\hat{Q}^{\mathrm{Weyl}}_{\mathcal{F}}(\eta,\omega)^*=\hat{Q}^{\mathrm{Weyl}}_{\mathcal{F}}(-\eta,-\omega),
\end{equation}
where the $*$ on the left hand side denotes the adjoint operator.
It is known that
the Weyl quantizer sends real functions into symmetric operators.
An immediate consequence of this  is the fact that any polynomial in $p$ and $q$ is sent to a symmetric polynomial of $\hat p,\hat q$. Different orderings are possible, for example we could order all $\hat p$ to the left, or the right, and they would be given by the quantizers 
\begin{align}
\hat{Q}^{\mathrm{L}}_{\mathcal{F}}(\eta,\omega)&=\e^{\ii \left(\eta \hat{q}- \omega \hat{p}\right)-\frac i2 \omega\eta}=\e^{\ii \eta \hat{q}}\e^{-  i\omega \hat{p}}&\nonumber\\
\hat{Q}^{\mathrm{R}}_{\mathcal{F}}(\eta,\omega)&=\e^{\ii \left(\eta \hat{q}- \omega \hat{p}\right)+\frac i2 \omega\eta}=\e^{-  i\omega \hat{p}}\e^{\ii \eta \hat{q}}&
\end{align}
The Fourier transforms of these operators are simple distribution valued operators:
\begin{align}
\hat{Q}^{\mathrm{L}}_{\mathcal{F}}(p,q)&=\delta(p-\hat p)\delta(q-\hat q)\nonumber\\
\hat{Q}^{\mathrm{R}}_{\mathcal{F}}(p,q)&=\delta(q-\hat q)
\delta(p-\hat p)
\end{align}
Intermediate orderings are also possible, and we define the $\gamma$ ordered quantizer, in Fourier transform, to be the operator:
\begin{equation}\label{ordergamma}
\hat{Q}^\gamma_{\mathcal{F}}(\eta,\omega)=\e^{\ii \left(\eta \hat{q}- \omega \hat{p}\right)+i\gamma \omega\eta}
%=\e^{\ii \eta \hat{q}}\e^{-  i\omega \hat{p}}\e^{\frac i2 \omega\eta}=\e^{-  i\omega \hat{p}}\e^{\ii \eta \hat{q}}\e^{-\frac i2 \omega\eta}.
\end{equation}

In the following similar definitions will be provided when dealing with quantizers and their Mellin transforms.
It is often desirable that a quantization scheme maps real functions into symmetric operators on $\mathcal{H}.$ Let us recall some well known results about quantization of classical observables.

%\begin{prop}\label{RSA}
The quantization scheme associated to the quantizer $\hat{Q}$ sends real functions into symmetric operators if and only if $\hat{Q}(p,q)$ is a symmetric function of $(p,q) \in  \mathbb{R}^{2N}.$
%\end{prop}
%\begin{proof}
%Let $f(x)$ be a function on $\mathbb{R}^{2N},$ let 
%\begin{equation}
%\hat{F}=\int_{\mathbb{R}^{2N}} \dd xf(x)\hat{Q}(x)
%\end{equation}
%be the operator associated to $f(x)$ by the quantization scheme defined by $\hat{Q}(x),$ then %the adjoint of $\hat{F}$ is 
%\begin{equation}
%\hat{F}^*=\int_{\mathbb{R}^{2N}} \dd x\overline{f}(x)\hat{Q}^*(x)
%\end{equation}
%if $f(x)$ is a real function, we have
%\begin{equation}
%\hat{F}^*=\int_{\mathbb{R}^{2N}} \dd xf(x)\hat{Q}^*(x)
%\end{equation}
%and $\hat{F}^*= \hat{F}$ if and only if $\hat{Q}(x)$ is \red{self-adjoint}.
%\end{proof}
In this case  we say that $\hat{Q}$ is a symmetric quantizer. 
%\begin{rema}In this paper we will not deal with domain issues for %operators over Hilbert spaces. In particular for every symmetric operator, we assume that a suitable choice of a \red{self-adjoint} extension is possible, and it has been made. \end{rema}
Moreover  the following holds:
\begin{equation}
\hat{Q}_{\mathcal{F}}((\eta,\omega))=\hat{Q}_{\mathcal{F}}^*(-\eta,-\omega)
\end{equation}
This follows from
\begin{equation}
\hat{F}=\int_{\mathbb{R}^{2N}} \dd q\dd p f(q,p) \hat{Q}(q,p)=\int_{\mathbb{R}^{2N}} \dd\eta\dd\omega \int_{\mathbb{R}^{2N}} \dd q\dd p \tilde{f}(\eta,\omega) \hat{Q}_{\mathcal{F}}(\eta,\omega)\e^{\ii (\eta q-\omega p)}
\end{equation}
and 
\begin{equation}
\hat{F}^*=\int_{\mathbb{R}^{2N}} \dd \eta\dd\omega \int_{\mathbb{R}^{2N}} \dd q\dd p \tilde{f}^*(\eta,\omega) \hat{Q}_{\mathcal{F}}^*(\eta,\omega)\e^{-\ii (\eta q-\omega p)}
\end{equation}
as well as the property that symmetric quantizers send real functions into symmetric operators.

A second remarkable property is that Weyl quantization  maps $L^2(\mathbb R^{2n})$ functions into Hilbert Schmidt, in a such way that the norm is preserved (up to a constant normalization factor). The Weyl map is therefore an \emph{isometry}. Moreover this gives an isometry.

In the next sections we will be also interested in defining a dequantizer for $\kappa$-Minkowski spacetime.
Losely speaking, a dequantization map is expected to do the inverse job of a quantization map, hence sending operators  into functions on a classical phase space. Let us recall Wigner's dequantization map, which is the inverse of Weyl's map~\eqref{WM1}:
\begin{align}\label{WM}
\Gamma^{\mathrm{Wigner}}:\; & Op(\mathcal{H}) \rightarrow  \mathcal{F}(\mathbb{R}^{2N})\nonumber
\\&
 \hat{A}\mapsto \mathrm{Tr}\left(\hat{A}\e^{\ii \left(\eta \hat{q}- \omega \hat{p}\right)}\right)
 \end{align}
for a suitable class of operators, which includes trace class
.
As before, we define the dequantizer as the following operator-valued function:
\begin{align}\label{QW}
\hat{D}^{\mathrm{Wigner}}_{\mathcal{F}}:\, & \mathbb{R}^{2N}\rightarrow Op(\mathcal{H})
\nonumber
\\&(\eta, \omega)\mapsto \e^{\ii \left(\eta \hat{q}- \omega \hat{p}\right)}.
\end{align}
\begin{rema} Like before, the subscript $\mathcal{F}$ suggests that we are using the Fourier transform: in particular the function resulting from~\eqref{WM} will be the Fourier transform of a function on the classical phase space.
\end{rema}
Like before, we define  $\hat{D}: \mathbb{R}^{2N}\rightarrow Op(\mathcal{H})$ be an operator-valued function, we call the following map
 \begin{eqnarray}
\Gamma: &Op(\mathcal{H})  \rightarrow \mathcal{F}(\mathbb{R}^{2N})
\\ &\hat{A}\mapsto \mathrm{Tr}\left(\hat{A}\hat{D}(x)\right)
\end{eqnarray}
the \emph{dequantization scheme associated to the dequantizer} $ \hat{D}(x).$

 A pair quantizers/dequantizer define a quantization/dequantization scheme. 
%
%
%We have the following:
%\begin{mydef}
%Consider two operator valued maps $\hat{Q}: \mathbb{R}^{2N}\rightarrow Op(\mathcal{H})$ and $\hat{D}: \mathbb{R}^{2N}\rightarrow Op(\mathcal{H})$ as quantizer and dequantizer respectively. Denote as $\Omega$ the quantization scheme associated to $\hat{Q}$ and as $\Gamma$ the dequantization scheme associated to $\hat{D}.$ Then if 
%\begin{equation}
%\Omega \circ \Gamma (\hat{A})= \hat{A}\quad \mathrm{and} \quad \Gamma \circ \Omega (f)=f
%\end{equation}
%for every $\hat{A} \in Op(\mathcal{H})$ and for every $f \in \mathcal{F}(\mathbb{R}^{2N})$, we say that the dequantizer $\hat{D}$ is the inverse of the quantizer $\hat{Q}$ and vice-versa.
%\end{mydef}
When no confusion arises we shall simply say that  $\hat{D}$ is the dequantizer of $\hat{Q}.$
Weyl's quantization map and Wigner dequantization are paired in this sense. It is a well known fact that the operator-valued function~\eqref{QW} works as its own inverse\footnote{Up to a complex conjugation, depending if we are in Fourier transform space or not. And whether factors of $2\pi$ are absorbed in the measure or not.} when used as dequantizer. This is not true for a generic ordering.

\section{$\kappa$-Minkowski spacetime and the Mellin transform.}

It has been noted in~\cite{Dabrowski:2010yk,Lizzi:2018qaf,Lizzi:2019wto} that the $\kappa$-Minkowski commutation relation~\eqref{kmcr} can be written  in polar coordinates as
\begin{equation} \label{commrt}
[t,r]=\ii r \ ; \ [t,f(\theta,\varphi)]=0
\end{equation}
where to ease notations we indicate $x_0$ by $t$. The polar coordinates $r,\theta$ and $\varphi$ have the usual meaning, and we have chosen units in which $c=\kappa=1$. We will see below that the expression of functions in polar coordinates is crucial for the construction of the quantizers based on Mellin transforms. In what follows $\hbar$ plays no role, since our aim is to quantize spacetime only, this justifies the fact that we have an operator corresponding to time.
The commutation relation~\eqref{commrt} indicates that $\hat  t$ acts as a dilation operator on the functions of $r$:
\begin{equation}\label{dilat}
\hat t\, \psi(r,\theta,\varphi)=i \left(r\del_r +\frac32\right) \psi(r,\theta,\varphi)
\end{equation}
where the $\frac32$ factor is needed to make the operator symmetric. The (improper) eigenfunctions of the dilation operator~\eqref{dilat} are
monomials of the kind $r^{-(\frac32+\ii\tau)}$,  they play for this case the role played by plane waves for the Weyl-Wigner case.

In the previous section $\hat p$ and $\hat q$ and their respective representations were related by a Fourier transform, the relation~\eqref{commrt}  suggests  $\hat t$ and $\hat r$ are connected by a \emph{Mellin} transform, which take into account the measure $r^2\dd r$ for the integration of the radial variable. 
Since the radial variables commute, the corresponding operators  are central, and we do not consider them further. 

The equivalent of the Fourier transform between eigenfunctions of the position and time operators are:
\begin{align}
   \psi(r)&=\frac1{\sqrt{2\pi}}\int_{-\infty}^{\infty}\dd\tau\, r^{-\frac32}\e^{-\ii\tau\log r} \widetilde\psi(\tau)\nonumber\\
    \widetilde \psi(\tau)&=\frac1{\sqrt{2\pi}}\int_0^\infty r^2 \dd r\, r^{-\frac32}\e^{\ii\tau\log r} \psi(r).
\end{align}
We will seek a quantization procedure based on the Mellin transform/antitransform,  in analogy of what we did in the previous section with the Fourier transform. From now on, by $\tilde{f}$   we will always denote the Mellin transform of a function $f$ rather than its Fourier transform. There are variants for the measure of the Mellin transform, we are in three dimensions, and therefore for us, following \cite{Lizzi:2018qaf,Lizzi:2019wto} the Mellin transform of a function $f$ is defined as\footnote{For ease of notation we will omit the integration limits, it is understood that radial coordinates are integrated from $0$ to $\infty$, while time like (and logarithmical variables below) go from $-\infty$ to $\infty$.} 
\begin{equation}
\tilde{f}(\rho,\tau)=\frac{1}{2\pi}\int r^2 \dd r \dd t\, r^{-3/2}\e^{\ii  \tau \log r} \rho^{-3/2} \e^{-\ii t \log \rho} f(r,t).
\end{equation}
 This transform has been chosen in order to be an isometry on square integrable functions whose transform may be defined. The inverse map is given by
\begin{equation}
f(r,t)= \frac{1}{2\pi} \int \rho^2\dd \rho \,  \dd\tau\, r^{-3/2}\e^{-\ii  \tau \log r} \rho^{-3/2} \e^{\ii t \log \rho} \tilde{f}(\rho,\tau). \label{inverse}
\end{equation}
We shall refer to the map above as "Mellin anti-transform" or just "anti-transform" when no confusion can arise.

Denote by $\overline{f}$ the complex conjugate of a function. Then $f(r,t)$ of the variables $r$ and~$t,$
 is real if and only if its Mellin transform $\tilde{f}(\rho,\tau)$ enjoys the following property:
\begin{equation}
\tilde{f}(\rho,\tau)=\rho^{-3} \overline{\tilde{f}}(\rho^{-1},-\tau)
\label{oldpro31}
\end{equation}
Indeed from~\eqref{inverse} we have
\begin{align}
\overline{f}(r,t)&=\int\dd \rho\dd\tau r^{-3/2}\e^{\ii  \tau \log r} \rho^{1/2} \e^{-\ii t \log \rho} \overline{\tilde{f}}(\rho,\tau)
\nonumber\\
&=
\int\dd \rho^{-1}\dd\tau r^{-3/2}\e^{-\ii  \tau \log r} \rho^{-5/2} \e^{\ii t \log \rho} \overline{\tilde{f}}(\rho^{-1},-\tau)
\nonumber\\
&=\int\dd \rho \dd\tau r^{-3/2}\e^{-\ii  \tau \log \rho} \rho^{1/2} \e^{\ii t \log \rho}\left( \rho^{-3}\overline{\tilde{f}}(\rho^{-1},-\tau)\right).
\end{align}
Hence $f(r,t)=\overline{f}(r,t)$ if and only if $
\tilde{f}(\rho,\tau)=\rho^{-3}\overline{\tilde{f}}(\rho^{-1},-\tau).
$

In the next sections we will also work with the logarithmic variables
\begin{equation}
x:= \log r \quad \mathrm{and} \quad \xi:= \log \rho 
\end{equation}
The integration measure for these variables are
\begin{equation} \label{measurexrho}
  r^2\dd r= \e^{3x}\dd x  \quad \mathrm{and} \quad \rho^2d\rho= \e^{3\xi}\dd \xi  \end{equation}
 and the $\kappa$-Minkowski commutation relations are now canonical.
 \begin{equation}
 [t,x]=i
 \end{equation} 
We have effectively reduced the commutation relations to a one dimensional case. In the $x$ variable the commutation relation is also canonical, with $x$ defined on the whole line, the crucial difference in our case is in the measure. 
  
The Mellin transform of a function is now written as 
\begin{equation}
\tilde{f}(\xi,\tau)=\frac1{2\pi}\int_{\mathbb{R}^{2}} \dd x \,\dd t \,M(x,\xi,t,\tau) f(x,t)
\end{equation}
where we have defined the integral kernel 
\begin{equation}
M(x,\xi,t,\tau)=\e^{\ii \tau x + \frac{3}{2}x}\e^{-\ii t \xi-\frac{3}{2}\xi}.
\end{equation}
The anti-transform is:
\begin{equation}
f(r,t)=\frac1{2\pi} \int_{\mathbb{R}^{2}}\dd x \,\dd t \,M(x,\xi,t,\tau)^{-1} \tilde{f}(\rho,\tau).
\end{equation}

To use the Mellin transform in a  quantization scheme we need to give a meaning to the transforms %of the constant function $f(r,t)=1$ as well as of 
of the coordinate functions $r$ and $t$, and in general to polynomials in these variables. 
This is also useful to check the commutation relations. Therefore, we will consider their Mellin transform in the distributional sense. If $g$ belongs to an appropriate test functions space,  and  $\{f_\epsilon\}$ is a sequence of  functions such that the sequence of regular distributions $\{T_{f_\epsilon}\}$ defined by
\begin{equation}
 T_{f_\epsilon}(g)=\int_{\mathbb{R}^{2}}\dd\mu  f_\epsilon g
\end{equation}
for $\epsilon\to 0$
converges in the weak-$*$ topology to $T_f,$ then if the limit of the sequence $\{T_{\tilde{f}_\epsilon}\}$ exists in the weak-$*$ topology. We denote it as the Mellin transform of $f$ in the sense of distributions (or just the Mellin transform of $f$ for short).
In the following examples we provide some distributional Mellin transforms that will turn useful later. 
\begin{exa}
Consider the sequence $f_\epsilon(x,t)= \e^{-\epsilon x^2}$ which converges point-wise (hence also weakly-$*$) to the constant function equal to $f(x,t)=1.$ Then we have
\begin{align}
\lim_{\epsilon \rightarrow 0} \tilde{f}_\epsilon(x,t)&= \frac{1}{2\pi}\lim_{\epsilon \rightarrow 0} \int \dd x \dd t\, \e^{\ii \tau x + \frac{3}{2}x}\e^{-\ii t \xi-\frac{3}{2}\xi} \e^{-\epsilon x^2}=\lim_{\epsilon \rightarrow 0} \frac{\delta(\xi)\e^{-\frac{(\frac{3}{2}i-\tau)^2}{4\epsilon}}}{2 \sqrt{\pi}\sqrt{\epsilon}}
\nonumber \\& =2\pi \delta(\xi) \delta\left(\tau- \frac{3}{2}i\right).
\end{align}
\end{exa}
\begin{exa} 
Let  $r_\epsilon(x,t)= \e^{x} \e^{-\epsilon x^2}$ which converges pointwise to the function $r=\e^{x}.$ We have
\begin{align}
\lim_{\epsilon \rightarrow 0} \tilde{r}_\epsilon(x,t)&= \frac{1}{2\pi}\lim_{\epsilon \rightarrow 0} \int \dd x \dd t 
\, \e^xe^{\ii \tau x + \frac{3}{2}x}\e^{-\ii t \xi-\frac{3}{2}\xi} \e^{-\epsilon x^2}=\lim_{\epsilon \rightarrow 0} \frac{\delta\left(\tau-\frac{5}{2}i\right)\e^{-\frac{3\xi}{2}-\frac{\xi^2}{\epsilon}}}{\sqrt{\epsilon}}
\nonumber\\& = 2 \pi \delta(\xi) \delta\left(\tau - \frac{5}{2}i\right).
\end{align}
\end{exa}
\begin{exa} 
Let $t_\epsilon(x,t)=t \e^{-\epsilon x^2}$ which converges pointwise to the function $t.$ We have
\begin{align}
\lim_{\epsilon \rightarrow 0} \tilde{t}_\epsilon(x,t)&= \frac{1}{2\pi}\lim_{\epsilon \rightarrow 0} \int \dd x \, \e^te^{\ii \tau x + \frac{3}{2}x}\e^{-\ii t \xi-\frac{3}{2}\xi} \e^{-\epsilon x^2}=\lim_{\epsilon \rightarrow 0} -\frac{\delta\left(\tau-\frac{3}{2}i\right)\partial_\xi \e^{-\frac{3\xi}{2}-\frac{\xi^2}{\epsilon}}}{\sqrt{\epsilon}}
\nonumber\\
& =- 2 \pi i \e^{-\frac{3}{2}\xi}\partial_\xi \delta(\xi) \delta\left(\tau - \frac{3}{2}i\right).
\end{align}
\end{exa}
\begin{exa}
From the examples above it is easy to show that for $f(r,t)=r^mt^n$ with $m,n>0$ the Mellin transform is given by
\begin{equation}
\tilde{f}(\rho,\tau)= 2\pi(-i)^n \e^{-\frac{3}{2}\xi}\partial^n_\xi \delta(\xi)\delta\left(\tau-\frac{3+2m}{2}\right).
\end{equation}
\end{exa} 
\section{Quantizers and Dequantizers}
In this section we build a quantization scheme  based on the Mellin transform.
In analogy to what is done for Weyl quantization, we start by defining the Mellin transformed version of a quantizer to be 
\begin{equation}
\int r^2 \dd r  \dd t \hat{Q}(r,t)f(r,t)=\int \rho^2 \dd \rho  \dd \tau \, \hat{Q}_{\mathcal{M}}(\rho,\tau)\tilde{f}(\rho,\tau)
\end{equation}
and  set the notation
\begin{equation}
\hat{Q}_{\mathcal{M}}(f):= \int \rho^2 \dd \rho   \dd \tau \, \hat{Q}_{\mathcal{M}}(\rho,\tau)\tilde{f}(\rho,\tau).
\end{equation}  
Let us start by considering the quantizer ``with position on the left" in the physical space: it is given by:
\begin{equation} \label{qpl}
\hat{Q}_L(r,t)=\frac{\delta(\hat{r}-r)}{r^2}\delta(\hat{t}-t)
\end{equation}
An explicit calculation (details in the appendix, equation~\eqref{detail1}) shows that
\begin{equation} \label{Q(f)}
\hat{Q}(f)= 
\int \dd\xi  \dd \tau \; \hat{Q}_{\mathcal{M}_L}(\xi,\tau)\tilde{f}(\xi,\tau)
\end{equation}
where we have defined 
\begin{equation}\label{QLM}
\hat{Q}_{\mathcal{M}_L}(\xi,\tau):=\frac{1}{2\pi}\int \dd\xi  \dd \tau  M(x,t,\xi,\tau)^{-1} \hat{Q}_L(x,t)= \frac{1}{2\pi}\e^{\left(-i\tau-\frac{3}{2}\right)\hat x}\e^{\left(i\hat{t}+\frac{3}{2}\right)\xi}.
\end{equation}
Using  the Baker-Campbell-Haursdorff formula
%\begin{equation}
%e^Ae^B=e^{A+B+\frac{1}{2}[A,B]}
%\end{equation}
this quantizer can also be written in the more compact form
\begin{equation}
\hat{Q}_{\mathcal{M}_L}(\xi,\tau)=\frac{1}{2\pi}\e^{\left(-i\tau-\frac{3}{2}\right)\hat{x}+\left(i\hat{t}+\frac{3}{2}\right)\xi-\frac{1}{2}\xi\left(i\tau+\frac{3}{2}\right)}.
\end{equation}
It is easy to verify that, had we started in equation~\eqref{qpl}  with the "the position on the right" quantizer
\begin{equation}
\hat{Q}_R(r,t)=\delta(\hat{t}-t)\frac{\delta(\hat{r}-r)}{r^2}
\end{equation}
we would have ended up with the following expression for its Mellin transform
\begin{equation}
\hat{Q}_{\mathcal{M}_R}(\xi,\tau)=\frac{1}{2\pi}\e^{\left(-i\tau-\frac{3}{2}\right)\hat x+\left(i\hat{t}+\frac{3}{2}\right)\xi+\frac{1}{2}\xi\left(i\tau+\frac{3}{2}\right)}.
\end{equation}
More generally we can consider a one-parameter family of quantizers of the form
\begin{equation}\label{qgamma}
\hat{Q}_{\mathcal{M}_\gamma}(\xi,\tau)=\frac{1}{2\pi}\e^{\left(-i\tau-\frac{3}{2}\right)\hat{x}+\left(i\hat{t}+\frac{3}{2}\right)\xi+\frac{\gamma}{2}\xi\left(-i\tau-\frac{3}{2}\right)}.
\end{equation}
with $\gamma \in \mathbb{C}.$
The first obvious property that our quantizers should verify is the compatibility between classical and quantum coordinates, i.e.\ we should have:
\begin{equation}
\hat{Q}_{\mathcal{M}_\gamma}(\tilde{r})=\hat{r} \quad \mathrm{and} \quad \hat{Q}_{\mathcal{M}_\gamma}(\tilde{t})=\hat{t}.
\end{equation}
As far as the quantization of $r$ in concerned we have 
\begin{equation}
\hat{Q}_{\mathcal{M}_\gamma}(\tilde{r})=\frac{1}{2\pi}\int \dd \tau  \dd \xi \e^{\left(-i\tau-\frac{3}{2}\right)\hat{x}+\left(i\hat{t}+\frac{3}{2}\right)\xi+\frac{\gamma}{2}\xi\left(-i\tau-\frac{3}{2}\right)}2 \pi \delta(\xi) \delta\left(\tau - \frac{5}{2}i\right)=\e^{\hat{x}}=\hat{r}
\end{equation}
while for the quantization of $t$ we have
\begin{equation}
\hat{Q}_{\mathcal{M}_\gamma}(\tilde{t})=\frac{1}{2\pi} \dd \tau  \int \dd \xi \e^{\left(-i\tau-\frac{3}{2}\right)\hat{x}+\left(i\hat{t}+\frac{3}{2}\right)\xi+\frac{\gamma}{2}\xi\left(-i\tau-\frac{3}{2}\right)}\left[- 2 \pi i \e^{-\frac{3}{2}\xi}\partial_\xi \delta(\xi) \delta\left(\tau - \frac{3}{2}i\right)\right]=\hat{t}.
\end{equation}
This is independent from the value of $\gamma$, which shows that any $\gamma$ will give an acceptable quantizer. We now further characterise them.

If $\hat{Q}(x,t)$ is symmetric, then  $\hat{Q}_{\mathcal{M}}(\xi,\tau)$ has the following property.
\begin{equation}
\hat{Q}_{\mathcal{M}}^*(-\xi,-\tau)=e^{-3\xi}\hat{Q}_{\mathcal{M}}(\xi,\tau).
\end{equation}

If $\hat{Q}(r,t)$ is symmetric, it maps real function $f(r,t)$ into a symmetric operator.  Combining the definition of the Mellin transform of an operator and~\eqref{oldpro31},  for $f(x)$ real we have 
\begin{equation}
\hat{F}=\int \dd\xi\dd\tau\, \hat{Q}_{\mathcal{M}}(\xi,\tau) f(\xi,\tau)= \int\dd\xi\dd\tau\, \hat{Q}_{\mathcal{M}}(\xi,\tau)e^{-3\xi} \overline{f}(-\xi,-\tau)
\end{equation}
and 
\begin{equation}
\hat{F}^*= \int \dd\xi\dd\tau\,\hat{Q}_{\mathcal{M}}^*(\xi,\tau) \overline{f}(\xi,\tau)=\int\dd\xi\dd\tau\,  \hat{Q}_{\mathcal{M}}^*(-\xi,-\tau) \overline{f}(-\xi,-\tau)
\end{equation}
which characterizes the Mellin transform of the symmetric quantizer.
For the family of quantizers of the form~\eqref{qgamma} it is easily seen that only the choice $\gamma=0$ meets this conditions.
% since we have
%\begin{align}
%\hat{Q}_{\mathcal{M}_0}(\xi,\tau)&=\frac{1}{2\pi}\e^{\left(-i\tau-\frac{3}{2}\right)\hat{x}+\left(i\hat{t}+\frac{3}{2}\right)\xi}
%\nonumber\\e^{3\xi}\hat{Q}_{\mathcal{M}_0}^*(-\xi,-\tau)&=\frac{1}{2\pi}\e^{3\xi+\left(-i\tau-\frac{3}{2}\right)\hat{x}+\left(i\hat{t}-\frac{3}{2}\right)\xi}=\frac{1}{2\pi}\e^{\left(-i\tau-\frac{3}{2}\right)\hat{x}+\left(i\hat{t}+\frac{3}{2}\right)\xi}.
%\end{align}
This is in analogy with what we found in the Weyl-Wigner case earlier.

Another important analogy is given by the fact that the image of a function $f\in L^2(\mathbb{R}^2)$ by the quantizer $\hat{Q}_{\mathcal{M}_0}$ has a finite Hilbert-Schmidt norm. To see we represent the operators on functions of $x$, and denote $ \ket x (\bra x)$ the improper ket (bra) eigenfunctions of the operator $\hat{x}$. The normalisation must take into account the measure~\eqref{measurexrho}:
\begin{equation}
\int \dd x\, \e^{3x }\ket x\! \bra x=1
\end{equation}
we then have
\begin{equation}
\mathrm{Tr}\left(F^*F \right)= 
\int d x \, \e^{3x} \bra{x} 
\int \dd \xi'  \dd\tau'\overline{\tilde{f}(\xi',\tau')}\ \hat{Q}_{\mathcal{M}_\gamma}^*(\xi',\tau')
\int \dd \xi  \dd\tau\tilde{f}(\xi,\tau) \hat{Q}_{\mathcal{M}_\gamma}(\xi,\tau)
\ket{x}
\end{equation}
Let us express the expression above as follows: %\color{red}{\large\bf Tutto quanto segue \`e \emph{sbagliato!} Si integra in $\rho$ funzioni di $\xi$, senza tener conto della misura, mancano primi eccetera. A me non sembra che la mappa valga per  tutti i valori di $\gamma$, e va fatto con il quantizer arbitrario, altrimenti che stiamo facendo? Per favore Alessandro e/o Mattia riscrivete in forma corretta. I dettagli secondo me devono andare in appendice, come per la~\eqref{Q(f)}.}
{\begin{align}
\mathrm{Tr}\left(F^*F \right)=&\int \dd x \, \e^{3x} \bra{x} 
\int \dd\xi' \dd\tau'\overline{\tilde{f}(\xi',\tau')}\ \e^{\left(i \tau' - \frac{3}{2}\right)\hat x}\e^{\left(-i \hat t + \frac{3}{2}\right)\xi'}\e^{\left(-i \tau' + \frac{3}{2}\right)\xi'}\e^{\frac{\gamma}{2}\xi'\left(-i\tau-\frac{3}{2}\right)}
\nonumber\\&
\int \dd \xi \dd\tau\tilde{f}(\xi,\tau)\e^{\left(i \hat t + \frac{3}{2}\right)\xi} \e^{\left(-i \tau - \frac{3}{2}\right)\hat x}\e^{\left(i \tau + \frac{3}{2}\right)\xi}\ket{x}\e^{\frac{\bar \gamma}{2}\xi'\left(+i\tau'-\frac{3}{2}\right)}
\nonumber\\
=&\int\dd t \dd  x \, \e^{3x} \bra{x} 
\int  \dd \xi' \dd\tau'\overline{\tilde{f}(\xi',\tau')}\ \e^{\left(i \tau' - \frac{3}{2}\right)x}\e^{\left(i \hat t + \frac{3}{2}\right)\xi'}\e^{\left(-i \tau' + \frac{3}{2}\right)\xi'}\ket{t}\bra t
\nonumber\\
&\int  \dd  \xi \dd\tau\tilde{f}(\xi,\tau) \e^{\left(-i \hat t + \frac{3}{2}\right)\xi} \e^{\left(-i \tau - \frac{3}{2}\right)\hat x }\e^{\left(i \tau + \frac{3}{2}\right)\xi}\ket{x}
\e^{\frac{\gamma}{2}\xi\left(-i\tau-\frac{3}{2}\right)+\frac{\bar \gamma}{2}\xi'\left(+i\tau'-\frac{3}{2}\right)}
\nonumber\\
=&\int\dd t \dd  x \, \e^{3x} \bra{x} 
\int \dd \xi' \dd\tau'\overline{\tilde{f}(\xi',\tau')}\ \e^{\left(i \tau' - \frac{3}{2}\right)x}\e^{\left(-i t + \frac{3}{2}\right)\xi'}\e^{\left(-i \tau' + \frac{3}{2}\right)\xi'}\ket{t}\bra t
\nonumber\\
&\int \dd  \xi \dd\tau\tilde{f}(\xi,\tau)\e^{\left(-i \tau - \frac{3}{2}\right)x }\e^{\left(i t+ \frac{3}{2}\right)\xi}\e^{\left(i \tau + \frac{3}{2}\right)\xi}\ket{x}
\e^{\frac{\gamma}{2}\xi\left(-i\tau-\frac{3}{2}\right)+\frac{\bar \gamma}{2}\xi'\left(+i\tau'-\frac{3}{2}\right)}
\nonumber\\
=&\int\dd t \dd x  
\int \dd\xi' \dd\tau'\ \e^{\left(i \tau' - \frac{3}{2}\right)x}\e^{\left(-i t + \frac{3}{2}\right)\xi'}\e^{\left(-i \tau' + \frac{3}{2}\right)\xi'}
\nonumber\\&\int  d \xi   \dd\tau \e^{\left(-i \tau - \frac{3}{2}\right)x }\e^{\left(i t+ \frac{3}{2}\right)\xi}\e^{\left(i \tau + \frac{3}{2}\right)\xi}\overline{\tilde{f}(\xi',\tau')}\tilde{f}(\xi,\tau)
\e^{\frac{\gamma}{2}\xi\left(-i\tau-\frac{3}{2}\right)+\frac{\bar \gamma}{2}\xi'\left(+i\tau'-\frac{3}{2}\right)}.
\end{align}}
 We integrate in $\dd t$ to obtain
\begin{align}
&\int \dd x  
\int \dd\xi' \dd\tau'\ \delta(\xi-\xi') \e^{\left(i \tau' - \frac{3}{2}\right)x}\e^{\frac{3}{2}\xi'}\e^{\left(-i \tau' + \frac{3}{2}\right)\xi'}\nonumber\\
&
\int  \dd \xi   \dd\tau \e^{\left(-i \tau - \frac{3}{2}\right)x }\e^{\frac{3}{2}\xi}\e^{\left(i \tau + \frac{3}{2}\right)\xi}\overline{\tilde{f}(\xi',\tau')}\tilde{f}(\xi,\tau) \e^{\frac{\gamma}{2}\xi\left(-i\tau-\frac{3}{2}\right)+\frac{\bar \gamma}{2}\xi'\left(+i\tau'-\frac{3}{2}\right)}
\end{align}
 In order to integrate in $\dd x$ we make the following change of variables 
 \begin{equation}
              i \tau' - \frac{3}{2}\mapsto i\alpha' \quad -i \tau - \frac{3}{2} \mapsto  - i\alpha 
\end{equation} 
After integrating in $x$ we  have
 \begin{align}
&\int \dd\xi' \dd\alpha'\ \delta(\alpha-\alpha')\delta(\xi-\xi')\e^{\frac{3}{2}\xi'}\e^{-i \alpha' \xi'}
\int  \dd \xi   \dd\alpha \; \e^{\frac{3}{2}\xi}\e^{i\alpha\xi}\overline{\tilde{f}(\xi',\tau')}\tilde{f}(\xi,\tau)
\e^{\frac{\gamma}{2}\xi\left(-i\alpha\right)+\frac{\bar \gamma}{2}\xi'\left(i\alpha\right)}.
\nonumber\\&= \int  \dd \xi   \dd\alpha \; \e^{3\xi}\,\overline{\tilde{f}(\xi,\alpha)}\tilde{f}(\xi,\alpha)
\e^{\Re \gamma\xi\alpha}.
\end{align}
We see that the integral equals the norm $\lVert f \rVert_{L_{2}}^2$ for $\gamma$ pure imaginary or zero. To summarize:
%\begin{prop}
Given a quantizer of the form~\eqref{qgamma}, only the choice $\gamma=0$ has the properties of sending real functions into symmetric operators and of being an isometry between $L^2(\mathbb{R}^2)$ and the space of Hilbert-Schmidt operators.
%            \color{red}{
%Let us focus on the $x$ integration
%\begin{align}
%\int \dd x\ \e^{3x} \bra{x}  \hat{Q}^*(\xi',\tau') %\hat{Q}(\xi',\tau')\ket{x}=& \int \dd x\  \e^{3x}
%\frac{1}{4\pi}\e^{-\ii\tau\xi} \e^{\ii\tau'\xi'} \bra{x} \e^{\left(\ii\tau-\frac{3}{2}\right)\hat{x}}\e^{-\ii\hat{t}\xi} 
%\nonumber\\&\int \dd t \ket{t}\bra{t}\e^{\ii\hat{t}\xi'} \e^{\left(\ii\tau'+\frac{3}{2}\right)\hat{x}}\ket{x}=\nonumber\\
%=&\frac{1}{4\pi}\e^{-\ii\tau\xi} \e^{\ii\tau'\xi'}\int \dd x \e^{3x} \braket{x|t}\braket{t|x}\e^{\ii(\tau-\tau')x-3x}
%\nonumber\\&
%\int \dd t \ \e^{\ii t(\xi'-\xi)}\nonumber\\
%=&\frac{1}{2\pi}\e^{-\ii\tau\xi} \e^{\ii\tau'\xi'}\delta(\xi-\xi')\int \dd x \ \e^{\ii(\tau-\tau')x-3x}\nonumber\\
%=&e^{-\ii\tau\xi} %\e^{\ii\tau'\xi'}\delta(\xi-\xi')\delta(\tau-\tau'+\ii 3).
%\end{align}
%Thus, we have 
%\begin{align}
%\mathrm{Tr}\left(F^*F \right)&=\int \dd\xi \ \ \dd\tau \int \dd \xi'\ \dd \tau' \overline{f(\xi,\tau)}f(\xi',\tau') \e^{-\ii\tau\xi} \e^{\ii\tau'\xi'}\delta(\xi-\xi')\delta(\tau-\tau'+\ii 3)\nonumber\\
%&=\int \dd\xi \ \ \dd\tau \e^{-3\xi}\overline{f(\xi,\tau)}f(\xi,\tau).
%\end{align}
%And finally:
%\begin{equation}
%\mathrm{Tr} (F^*F)= \int \dd\mu(\xi,\tau)\overline{f(\xi,\tau)}f(\xi,\tau)= \lVert f \rVert_{L_{2}}^2.
%\end{equation}
%\end{proof}
%\section{Wigner-Mellin maps.}

We now compute the dequantizers of the Weyl-Mellin maps studied in the previous section. The quantizer-dequantizer pair should respect the following relation:
\begin{equation}\label{rQD}
\mathrm{Tr}(\hat Q_{\mathcal{M}_{\gamma}}(\xi,\tau) \hat D_{\mathcal{M}_\gamma}(\xi',\tau'))=\e^{-3\xi}\delta\left(\xi-\xi'\right)\delta\left(\tau-\tau'\right).
\end{equation}
Based on our experience with the Weyl Wigner case we make the following ansatz:
\begin{equation}
\hat D_{\mathcal{M}_\gamma}(\xi,\tau)=N\e^{a(\xi,\tau)\hat x+ b (\xi,\tau)\hat t+ c(\xi,\tau)}
\end{equation}
where $N$ is a normalization constant and $c$ a generic function of $\xi$ and $\tau$.
The commutator of the exponents of $Q_{\mathcal{M}_\gamma}$ and $D_{\mathcal{M}_\gamma}$ is given by
\begin{align}
&\left[\left(-i\tau-\frac{3}{2}\right)\hat{x}+\left(i\hat{t}+\frac{3}{2}\right)\xi+\frac{\gamma}{2}\xi\left(-i\tau-\frac{3}{2}\right),a(\xi',\tau')\hat x+ b (\xi',\tau')\hat t+ c(\xi',\tau')\right]
\nonumber\\
&=-\tau b(\xi',\tau')+ i \frac{3}{2}b(\xi',\tau')-\xi a (\xi',\tau').
\end{align}
Moreover one has 
\begin{equation}
\mathrm{Tr}\left(\e^{\ii k\hat{x}+i p\hat{t}}\right)= 2\pi \delta(k)\delta(p)
\end{equation}
hence we find
\begin{align}
\mathrm{Tr}(\hat Q_{\mathcal{M}_{\gamma}}(\xi,\tau) \hat D_{\mathcal{M}_\gamma}(\xi',\tau'))
=& N \delta\left(-\tau+ i \frac{3}{2}-ia(\xi',\tau')\right)\delta\left(\xi-ib(\xi',\tau')\right)
\nonumber\\
&\e^{\frac{3}{2}\xi+\frac{\gamma}{2}\xi(-i\tau-\frac{3}{2})+c(\xi',\tau')-\tau \frac b2(\xi',\tau') +i \frac{3}{2}\frac b2(\xi',\tau')-\xi \frac a2(\xi',\tau')}.
\end{align}
In order to satisfy equation~\eqref{rQD} we must have
\begin{equation}
a(\xi',\tau')= i\tau' +\frac{3}{2}, \quad b(\xi',\tau')=-i\xi' \quad \mathrm{and} \quad c(\xi',\tau')=-\frac{\gamma}{2}\xi'\left(-i\tau'- \frac{3}{2}\right)-\frac{3}{2}\xi'.
\end{equation}
Finally we have the Wigner-Mellin dequantizers:
\begin{equation}\label{deqM}
\hat{D}_{\mathcal{M}_\gamma}(\xi,\tau)= \e^{\left(i\tau+\frac{3}{2}\right)\hat{x}+\left(-i\hat{t}-\frac{3}{2}\right)\xi+\frac{\gamma}{2}\xi\left(i\tau+\frac{3}{2}\right)}.
\end{equation}

We have seen the behaviour of the pair quantizer/dequantizer $\hat{Q}_{\mathcal{M}_\gamma}/\hat{D}_{\mathcal{M}_\gamma}$ for $\kappa$-Minkowski spaces closely mirrors that of the one for a canonical noncommutative space, in being an isometry between $L^2(\mathbb R^)$ and Hilbert-Schimdt spaces operators.One noteworthy difference in our case is that this result holds true for every value of the parameter $\gamma \in \mathbb{C}.$
\section{The $*$ product}
As for the Weyl case, a quantization map always define a deformed product among functions.
Let us recall the standard definition of  the Gr\"onewold-Moyal's $*$ product adapted to the case at hand.
\begin{mydef}
Given a pair $(\hat Q,\hat D)$ of a quantizer and a dequantizer, the $*-$product of two functions $f(r,t)$ and $g(r,t)$ is defined as
\begin{equation}
(f*g)(r,t)=\mathrm{Tr}(\hat{Q}(f)\hat{Q}(g)\hat{D}).
\end{equation}
\end{mydef}
In our case, we use the pair of quantizer-dequantizer $(Q_{\mathcal{M}_\gamma},D_{\mathcal{M}_\gamma})$ and we define the $*$ product as the Mellin anti-transform of the following expression.
\begin{equation}
\widetilde{(f*g)}(\rho,\tau)=\mathrm{Tr}(\hat{Q}_{\mathcal{M}_\gamma}(f)\hat{Q}_{\mathcal{M}_\gamma}(g)\hat{D}_{\mathcal{M}_\gamma}).
\end{equation}
As in the previous calculations, it is usually easier to use the $x$ and $\xi$ variables to perform the calculations.
\begin{exa}
The first check is to verify that $r*t-t*r=i r$. Thus, 
we  compute the $*$ product of $\hat{r}=e^{x}$ and $\hat{t}$. 
\begin{align}
\text{Tr}[\hat{r}\hat{t} \hat{D}_{\mathcal{M}_\gamma}]=&\int \dd x e^3 \bra{x} e^{x}\int \dd t \ket{t}\bra{t}\hat{t} e^{-i \hat{t}\xi}e^{(i\tau +\frac{3}{2})\hat{x}}\ket{x}e^{-\frac{3}{2}\xi}e^{{-\frac{\gamma+1}{2}}(i\tau+\frac{3}{2})\xi} \nonumber\\
&=e^{-\frac{3}{2}\xi}e^{{-\frac{\gamma+1}{2}}(i\tau+\frac{3}{2})\xi}\int \dd x e^{(i\tau+\frac{3}{2})x}\int \dd t\  t e^{-it\xi}\nonumber\\  
&=e^{-\frac{3}{2}\xi}e^{{-\frac{\gamma+1}{2}}(i\tau+\frac{3}{2})\xi} \ 2\pi \ \delta\left(\tau-i\dfrac{3}{2}\right)\ (-i)\partial_\xi\delta(\xi)
\end{align}
Similarly one gets
\begin{equation}
\text{Tr}[\hat{t} \hat{r}\hat{D}_{\mathcal{M}_\gamma}]=e^{-\frac{3}{2}\xi}e^{{-\frac{\gamma-1}{2}}(i\tau+\frac{3}{2})\xi} \ 2\pi \ \delta\left(\tau-i\frac{3}{2}\right)\ (-i)\partial_\xi\delta(\xi)
\end{equation}
and the $*$ product is given by
\begin{align}
r * t &= \frac{1}{2 \pi} \int \dd\xi \int \dd \tau e^{-(i\tau+\frac{3}{2})x}e^{(i t +\frac{3}{2})\xi}\text{Tr}[\hat{r}\hat{t} \hat{D}_{\mathcal{M}_\gamma}]\nonumber\\
&=\int \dd\xi \int \dd \tau \ e^{-(i\tau+\frac{3}{2})x}e^{i t \xi}e^{{-\frac{\gamma+1}{2}}(i\tau+\frac{3}{2})\xi} \ \ \delta\left(\tau-i\frac{3}{2}\right)\ (-i)\partial_\xi\delta(\xi)\nonumber\\
&=e^{x}\int \dd\xi e^{(i t-\frac{\gamma}{2}-1)\xi}(-i)\partial_\xi\delta(\xi)\nonumber\\
&= e^{x}\left(t-i\frac{\gamma+1}{2}\right)
\end{align}
thus
\begin{equation}
r * t = r\left(t-i\frac{\gamma}{2}\right)-i \frac{r}{2}, \
t * r = r\left(t-i\frac{\gamma}{2}\right)+i \frac{r}{2}
\end{equation}
which leads to the desired commutation rule
\begin{equation}
t * r -r * t= i \ r.
\end{equation}
\end{exa}
\begin{exa}
The generalization to the product of monomials is given by
\begin{align}
\text{Tr}[\hat{r}^n\hat{t}^m \hat{D}_{\mathcal{M}_\gamma}]=&\int \dd x e^3 \bra{x} e^{nx}\int \dd t \ket{t}\bra{t}\hat{t}^m e^{-i \hat{t}\xi}e^{(i\tau +\frac{3}{2})\hat{x}}\ket{x}e^{-\frac{3}{2}\xi}e^{{-\frac{\gamma+1}{2}}(i\tau+\frac{3}{2})\xi}
\nonumber\\
&=e^{-\frac{3}{2}\xi}e^{{-\frac{\gamma+1}{2}}(i\tau+\frac{3}{2})\xi}\int \dd x e^{(i\tau+\frac{3}{2}+n)x}\int \dd t\  t^m e^{-it\xi}  \nonumber\\
&=e^{-\frac{3}{2}\xi}e^{{-\frac{\gamma+1}{2}}(i\tau+\frac{3}{2})\xi} \ 2\pi \ \delta\left(\tau-i\left(\dfrac{3}{2}+n\right)\right)\ (-i)^m\partial^m_\xi\delta(\xi)
\end{align}
Similarly one gets
\begin{equation}
\text{Tr}[\hat{t}^m \hat{r}^nD]=e^{-\frac{3}{2}\xi}e^{{-\frac{\gamma-1}{2}}(i\tau+\frac{3}{2})\xi} \ 2\pi \ \delta\left(\tau-i\left(\frac{3}{2}+n\right)\right)\ (-i)^m\partial_\xi^m\delta(\xi)
\end{equation}
and the $*$ product is given by
\begin{align}
r^n * t^m &= \frac{1}{2 \pi} \int \dd\xi  \dd \tau e^{-(i\tau+\frac{3}{2})x}e^{(i t +\frac{3}{2})\xi}\text{Tr}[\hat{r}^n\hat{t}^m D]\nonumber\\
&=\int \dd\xi \dd \tau \ e^{-(i\tau+\frac{3}{2})x}e^{i t \xi}e^{{-\frac{\gamma+1}{2}}(i\tau+\frac{3}{2})\xi} \ \ \delta\left(\tau-i\left(\frac{3}{2}+n\right)\right)\ (-i)^m\partial^m_\xi\delta(\xi)\nonumber\\
&=e^{n x}\int \dd\xi e^{(i t-\frac{\gamma}{2}-1)n\xi}(-i)^m\partial^m_\xi\delta(\xi)\nonumber\\
&= e^{nx}\left(t-i\frac{\gamma+1}{2}\right)^m.
\end{align}
Thus, we have 
\begin{align}
r^n * t^m &= r^n\left(t-i\frac{\gamma+1}{2}\right)^m\nonumber\\
t^m * r^n &= r^n\left(t-i\frac{\gamma-1}{2}\right)^m.
\end{align}
\end{exa}
\begin{exa}
It is known that for the Gr\"onewold-Moyal case the product of two Gaussian functions is still a Gaussian. A similar property holds in our case for \emph{log-Gaussians}, i.e.\ functions of the kind
\begin{equation}
f(r,t)= \e^{-(a \log r+ bt)^2}=\e^{-(a x+ bt)^2}
\end{equation}
It is easy to see that the Mellin transform of this kind of functions is still a function of this kind in $\xi,\tau$ space, up to constants. In order not to make notations heavy we will ignore the constants and start directly in Mellin transforms for the $\gamma=0$ case.
\begin{equation}
\tilde{f}(\xi, \tau)=\e^{-\frac{\tau^2}{\sigma^2}-\frac{\xi^2}{\eta^2}+\frac{\xi}{2}\left(-i\tau-\frac{3}{2}\right)}
\end{equation}
and
\begin{equation}
\tilde{g}(\xi, \tau)=\e^{-\frac{\tau^2}{\sigma'^2}-\frac{\xi^2}{\eta'^2}+\frac{\xi}{2}\left(-i\tau-\frac{3}{2}\right)}
\end{equation}
We compute the Weyl-Mellin transform of $\tilde{f}$ to get the following 
\begin{align}
\hat{Q}_{\mathcal{M}_0}( \tilde{f})=&\frac{1}{2\pi}\int \dd\tau   \dd \xi \e^{\left(-i\tau-\frac{3}{2}\right)\hat{x}+\left(i\hat{t}+\frac{3}{2}\right)\xi}\tilde{f}(\xi,\tau)\nonumber\\=
%\\&\frac{1}{2\pi}\int d%%\tau \int  d \xi %e^{\left(-\frac{\tau^2}{\sigma}-i\tau%\hat{x}-\frac{3}{2}\hat{x}\right)+\left(-%%\frac{\xi^2}{\eta}+i\hat{t}\xi\right)}=
&\frac{1}{2\pi}\int \dd\tau  \dd \xi \e^{\left(-\frac{\tau^2}{\sigma}-i\tau\hat{x}-\frac{3}{2}\hat{x}\right)}\e^{\left(-\frac{\xi^2}{\eta}+i\hat{t}\xi\right)}
\nonumber\\=&\frac{\sqrt{\sigma \eta}}{2}\e^{-\frac{\hat t^2}{4}\eta}\e^{-\frac{\hat{x}^2}{4}\sigma-\frac{3}{2}\hat x}
\end{align}
Hence the Mellin quantization of a Gaussian function is again a Gaussian function taking values in $Op(\mathcal{H}).$ We find the same result for $\tilde{g}(\xi,\tau),$ hence to compute the $*$ we have to compute the following trace:
\begin{align}\label{extr}
\tilde{f}*\tilde{g}(\xi,\tau)=& \frac{\sqrt{\sigma\eta\sigma'\eta'}}{4}\mathrm{Tr}\left(e^{-\frac{\hat t^2}{4}\eta}\e^{-\frac{\hat{x}^2}{4}\sigma-\frac{3}{2}\hat x}\e^{-\frac{\hat t^2}{4}\eta'}\e^{-\frac{\hat{x}^2}{4}\sigma'-\frac{3}{2}\hat x}\e^{\left(i\tau+\frac{3}{2}\right)\hat x+ \left(-i\hat t-\frac{3}{2}\right)\xi}\right)
\nonumber\\=& \int\bra t \e^{-\frac{\hat t^2}{4}\eta}\e^{-\frac{\hat{x}^2}{4}\sigma-\frac{3}{2}\hat x}\e^{-\frac{\hat t^2}{4}\eta'}\e^{-\frac{\hat{x}^2}{4}\sigma'-\frac{3}{2}\hat x}\e^{\left(i\tau+\frac{3}{2}\right)\hat x+ \left(-i\hat t-\frac{3}{2}\right)\xi} \ket t 
\end{align}
A calculation (detailed in the appendix), shows that
\begin{equation}\label{loggausprod}
\tilde{f}*\tilde{g}(\xi,\tau)=4\pi^2\sqrt{\frac{1}{\sigma'\eta'(4-\sigma\eta)}}\e^{\frac{9}{\sigma^2}}\e^{-\frac{\xi^2 \sigma}{4-\sigma \eta}-\frac{\xi}{2}\left(-i\tau+\frac{3}{2}\right)}\e^{-\frac{\tau^2}{\sigma'}}
\end{equation}
\end{exa}
\begin{exa}
We want to discuss the composition law of two ``plane waves'', one in the time direction and the other along a space direction, say the vertical axis. In our frameworks, this will be given by the $*$ product of $\e^{\ii a \hat{t}}$ and $\e^{\ii b \hat{r} \cos\theta}$.
 First of all, we compute
\begin{align} \label{{TrPW}}
\mathrm{Tr}[\e^{\ii a \hat{t}}\e^{\ii b\cos\theta \e^{\hat{x}} } \hat{D}_{\mathcal{M}_\gamma}]=&\int \dd t  \bra{t} \e^{\ii a t}\int \dd x\  \e^{3x}\ket{x}\bra{x}\e^{\ii b\cos\theta \e^{\hat{x}} } \e^{(i\tau +\frac{3}{2})\hat{x}}\e^{-i \hat{t}\xi}\ket{t}\e^{-\frac{3}{2}\xi}\e^{{-\frac{\gamma-1}{2}}(i\tau+\frac{3}{2})\xi} \nonumber\\  
&=\e^{-\frac{3}{2}\xi}\e^{{-\frac{\gamma-1}{2}}(i\tau+\frac{3}{2})\xi}\int \dd x\  \e^{\ii b\cos\theta \e^{x} } \e^{(i\tau +\frac{3}{2})x}\int \dd t\ \e^{\ii t (a-\xi)}\nonumber\\
&= \sqrt{2 \pi} \e^{-\frac{3}{2}\xi}\e^{{-\frac{\gamma-1}{2}}(i\tau+\frac{3}{2})\xi} \ \delta(\xi-a) \int \dd x\  \e^{\ii b\cos\theta \e^{x} } \e^{(i\tau +\frac{3}{2})x}
\end{align}
The $*-$product is obtained taking the anti transform 
\begin{align}
\e^{\ii a t}*\e^{\ii b\cos\theta \e^x } &= \frac{1}{2 \pi} \int \dd\xi \int \dd \tau \e^{-(i\tau+\frac{3}{2})x}\e^{(i t +\frac{3}{2})\xi}\mathrm{Tr}[\e^{\ii a \hat{t}}\e^{\ii b\cos\theta \e^{\hat{x}} } \hat{D}_{\mathcal{M}_\gamma}]\nonumber\\
&=\frac{1}{2\pi}\int \dd\xi \int \dd \tau \ \e^{-(i\tau+\frac{3}{2})x}\e^{i t \xi}\e^{{-\frac{\gamma-1}{2}}(i\tau+\frac{3}{2})\xi} \ \sqrt{2\pi} \delta(\xi-a) \int \dd x'\  \e^{\ii b\cos\theta \e^{x'} } \e^{(i\tau +\frac{3}{2})x'}\nonumber\\
&= \frac{1}{\sqrt{2\pi}}\e^{i t a}\e^{{- \frac{3}{2}a\frac{\gamma-1}{2}}}\int \dd x' \e^{\ii b\cos\theta \e^{x'} }\e^{\frac{3}{2}(x'-x)}\ \int \dd \tau \ \e^{i\tau\left(x'- x-a\frac{\gamma-1}{2}\right)} \nonumber\\
&= \frac{1}{\sqrt{2\pi}} \e^{i t a}\e^{{- \frac{3}{2}a\frac{\gamma-1}{2}}}\int \dd x' \e^{\ii b\cos\theta \e^{x'} }\e^{\frac{3}{2}(x'-x)}\  \ \sqrt{2 \pi} \delta \left(x'- x-a\frac{\gamma-1}{2}\right)\\ \nonumber
&=\e^{i t a}\e^{\ii b\cos\theta\  \text{Exp}\left(x+a\frac{\gamma-1}{2}\right)}= \e^{i t a}\e^{\ii r b\cos\theta\  \text{Exp}\left(a\frac{\gamma-1}{2}\right)}
\end{align}
Thus, we have 
\begin{align}
\e^{\ii a t}*\e^{\ii r b\cos\theta }&=  \e^{i t a}\e^{\ii r b\cos\theta\  \text{Exp}\left(a\frac{\gamma-1}{2}\right)}\nonumber\\
\e^{\ii r b\cos\theta }*\e^{\ii a t}&=\  \e^{i t a}\e^{\ii r b\cos\theta\  \text{Exp}\left(a\frac{\gamma+1}{2}\right)}.
\end{align}
\end{exa}

\section{Conclusion and Outlook}
We have shown how the Mellin transform for $\kappa$-Minkowski defines a quantization procedure giving arbitrary orderings of operators corresponding to the radial and time coordinates. We also built the corresponding maps from operators to functions, and the relative dequantizers. For a particular ordering we have shown that real functions, i.e.\ classical oservables, go into symmetric operators, i.e.\ quantum observables.
This is on par with what happens for the usual $p,q$ quantization. The role that in the latter quantization scheme is played by $\hbar$, which acts as a deformation parameter, is here played by $1/\kappa$. A real challenge would be to have both constants present. This is not going to be immediate. It is known that a time operator, in the usual quantization, is problematic. But the use of function of time may be well defined~\cite{Aniello:2016nvp}, and this could lead to a quantization based on $*$ product deformations, and possible lend itself to matrix simulations and possible fuzzy versions along the lines of the reviews~\cite{Lizzi:2014pwa,DAndrea:2013rix}

We also built the $*$ product implicitly defined by our quantization-dequantization scheme. We have given some examples here, and checked the most relevant properties. All properties of the product, and how it might help in the dynamical quantization of fields on $\kappa$-Minkowski space is an important issue with mathematical and physics ramifications, which deserves further scrutiny. We hope to come back to it in the future. 

\appendix

\section{Details of some calculations}

The chain of equalities leading to equation~\eqref{Q(f)} is the following:
\begin{align} \label{detail1}
\hat{Q}(f)&= \int \e^{3x} dx  \int \dd t \,\hat{Q}_L(r,t) f(r,t) 
\nonumber\\& = \int \e^{3x} dx  \int \dd t \,\delta(\hat{x}-x)e^{-3x} \delta(\hat{t}-t) f(r,t)
\nonumber\\& = \int dx  \int \dd t \,
\hat{Q}_L(x,t)f(x,t)
\nonumber\\& = \frac{1}{2\pi}\int dx  \int \dd t \,
\hat{Q}_L(x,t)\int d\xi  \int  d \tau  M(x,t,\xi,\tau)^{-1}\tilde{f}(\xi,\tau)
\nonumber\\& = 
\int d\xi \int d \tau \; \hat{Q}_{\mathcal{M}_L}(\xi,\tau)\tilde{f}(\xi,\tau)
\end{align}
Let us fix the following short hand notation for the decompositions of the identity:
\begin{equation}
1_{\hat{t}}:=
\int\ket t \bra t \quad \mathrm{and} \quad 1_{\hat{x}}:=
\int\e^{3x}\ket x \bra x
\end{equation} The chain of equalities leading to equation~\eqref{loggausprod} is the following:
\begin{align}
&\int\dd t \bra t \e^{-\frac{\hat t^2}{4}\eta}1_{\hat{x}}\e^{-\frac{\hat{x}^2}{4}\sigma-\frac{3}{2}\hat x}1_{\hat{t'}}\e^{-\frac{\hat t^2}{4}\eta'}1_{\hat{x'}}\e^{-\frac{\hat{x}^2}{4}\sigma'-\frac{3}{2}\hat x}\e^{\left(i\tau+\frac{3}{2}\right)\hat x}\e^{\left(-i\hat t-\frac{3}{2}\right)\xi} \e^{-\frac{\xi}{2}\left(-i\tau-\frac{3}{2}\right)} \ket t 
\nonumber\\=&\int\bra t \e^{-\frac{ t^2}{4}\eta}1_{\hat{x}}\e^{-\frac{x^2}{4}\sigma-\frac{3}{2}x}1_{\hat{t'}}\e^{-\frac{t'^2}{4}\eta'}1_{\hat{x'}}\e^{-\frac{x'^2}{4}\sigma'-\frac{3}{2}x'}\e^{\left(-i\tau+\frac{3}{2}\right)x'}\e^{\left(-i t-\frac{3}{2}\right)\xi} \e^{-\frac{\xi}{2}\left(-i\tau-\frac{3}{2}\right)} \ket t
\nonumber\\
=&\int\bra t \e^{-\frac{ t^2}{4}\eta}1_{\hat{x}}\e^{-\frac{x^2}{4}\sigma-\frac{3}{2}x}\int\dd t' \ket {t'} \e^{-\frac{t'^2}{4}\eta'}\int \dd x'\bra {t'} x'\rangle \;e^{-\frac{x'^2}{4}\sigma'+(-i\tau+3)x'}\e^{\left(-i t-\frac{3}{2}\right)\xi} \e^{-\frac{\xi}{2}\left(-i\tau-\frac{3}{2}\right)} \langle x' \ket t 
\nonumber\\=&e^{-\frac{\xi}{2}\left(-i\tau-\frac{3}{2}\right)}\int\e^{\ii t}\e^{\left(-i t-\frac{3}{2}\right)\xi}\bra t \e^{-\frac{ t^2}{4}\eta}1_{\hat{x}}\e^{-\frac{x^2}{4}\sigma-\frac{3}{2}x}\int\dd t' \ket {t'}\e^{-\frac{t'^2}{4}\eta'}\int \dd x' \;e^{-\frac{x'^2}{4}\sigma'+i\tau x'}\e^{-\ii t'}
\nonumber
\end{align}
\begin{align}
=&\sqrt{\frac{4\pi}{\sigma'}}\e^{-\frac{\xi}{2}\left(-i\tau-\frac{3}{2}\right)}\e^{-\frac{\tau^2}{\sigma'}}\int\e^{\ii t}\e^{\left(-i t-\frac{3}{2}\right)\xi}\bra t \e^{-\frac{ t^2}{4}\eta}1_{\hat{x}}\e^{-\frac{x^2}{4}\sigma-\frac{3}{2}x}\int\dd t' \ket {t'}\e^{-\frac{t'^2}{4}\eta'}\e^{-\ii t'}\nonumber
\\=&\sqrt{\frac{4\pi}{\sigma'}}\e^{-\frac{\xi}{2}\left(-i\tau-\frac{3}{2}\right)}\e^{-\frac{\tau^2}{\sigma'}}\int\dd t \; \e^{-\frac{t^2}{\sigma}}\e^{\ii t}\e^{\left(-i t-\frac{3}{2}\right)\xi} \e^{-\frac{ t^2}{4}\eta}\int \dd x \;\bra t x\rangle  \e^{-\frac{x^2}{4}\sigma}\int\dd t'  \e^{-\frac{t'^2}{4}\eta'} \nonumber\\=&4\pi\sqrt{\frac{1}{\sigma'\eta'}}\e^{-\frac{\xi}{2}\left(-i\tau+\frac{3}{2}\right)}\e^{-\frac{\tau^2}{\sigma'}}\int\dd t \; \e^{-\frac{t^2}{\sigma}}\e^{-\ii  t\xi} \e^{-\frac{ t^2}{4}\eta}\int \dd x \; \e^{-\frac{x^2}{4}\sigma-\frac{3}{2}x }\nonumber\\=&4\pi\sqrt{\frac{\pi}{4\sigma\sigma'\eta'}}\e^{\frac{9}{\sigma^2}}\e^{-\frac{\xi}{2}\left(-i\tau+\frac{3}{2}\right)}\e^{-\frac{\tau^2}{\sigma'}}\int\dd t \; \e^{-t^2\left(\frac{4-\sigma\eta}{4\sigma}\right)}\e^{-\ii  t\xi}\nonumber\\=&4\pi^2\sqrt{\frac{1}{\sigma'\eta'(4-\sigma\eta)}}\e^{\frac{9}{\sigma^2}}\e^{-\frac{\xi^2 \sigma}{4-\sigma \eta}-\frac{\xi}{2}\left(-i\tau+\frac{3}{2}\right)}\e^{-\frac{\tau^2}{\sigma'}}.
\end{align}

\subsubsection*{Acknowledgments}
We wish to thank Alessandro Zampini for discussions. F. L. and M.M. acknowledge support from the INFN Iniziativa Specifica GeoSymQFT, F.L. the Spanish MINECO underProject No. MDM-2014-0369 of ICCUB (Unidad de Excelencia `Maria de Maeztu'), Grant No. FPA2016-76005-C2-1-P.  67985840. A.C. aknowledges support from GA\u{C}R project 20-17488Y and RVO: 67985840.
FM thanks the Action CA18108 QG-MM from the European Cooperation in Science and Technology (COST) and the Foundational Questions Institute (FQXi).

%\newpage

\providecommand{\href}[2]{#2}\begingroup\raggedright\endgroup

\end{document}